# A Model for Software Contexts

Diana Kirk and Stephen G. MacDonell
*SERL, School of Computing & Mathematical Sciences, Auckland University of Technology,
Private Bag 92006, Auckland 1142, New Zealand*

**Abstract**

*It is widely acknowledged by researchers and practitioners that software development methodologies are generally adapted to suit specific project contexts. Research into practices-as-implemented has been fragmented and has tended to focus either on the strength of adherence to a specific methodology or on how the efficacy of specific practices is affected by contextual factors. We submit the need for a more holistic, integrated approach to investigating context-related best practice. We propose a six-dimensional model of the problem-space, with dimensions organisational drivers (why), space and time (where), culture (who), product life-cycle stage (when), product constraints (what) and engagement constraints (how). We test our model by using it to describe and explain a reported implementation study. Our contributions are a novel approach to understanding situated software practices and a preliminary model for software contexts.*

## 1. INTRODUCTION

For many years, the established wisdom relating to software development projects was that a specific methodology must be selected and, once selected, must be adhered to without alteration. This viewpoint was based on the notion that "development models are best regarded as a coherent set of practices, some of which are required to balance the performance trade-offs arising from the use (or absence) of others" (Cusumano et al., 2003). As time has passed, it has become clear that this ideal has seldom been reached in practice (Avison and Pries-Heje, 2008; Bajec et al., 2007; Hansson et al., 2009; MacCormack et al., 2012; Petersen and Wohlin, 2009; de Azevedo Santos et al., 2011; Turner et al., 2010). Research indicates that, rather than adopt and adhere to a specific methodology, many organisations adapt practices from several approaches, often at the level of the individual project. As an example, as agile approaches have become more established, deficiencies have been exposed, leading to either contextualisation (Hoda et al., 2010) or to amalgamation with other paradigms, for example, the 'lean' paradigm (Wang et al., 2012). Mac- Cormack el al. suggest that firms must deploy different business processes according to business context and that applying a uniform 'best practice' approach results in missed opportunities (MacCormack et al., 2012). The emergence of global software development has raised a number of issues representative of this paradigm and researchers suggest current practices must be extended or modified (Richardson et al., 2012; ul Haq et al., 2011). The software-as- a-service paradigm has raised new possibilities for evaluation of user behaviour and new kinds of approach are required as current methodologies do not satisfy the need for these new opportunities (Stucken-berg and Heinzl, 2010). The traditional viewpoint has thus been superseded by one of acceptance that tailoring according to project-specific contexts is both necessary and unavoidable.

The idea that practice and context are inherently intertwined has precedence in other areas of research. Studies in organisational and management sciences have applied theories of technology appropriation and task-technology fit to understand the adaptation of technologies by individuals and groups (Fidock and Carroll, 2006; Fuller and Dennis, 2009). 'Practices theory' is based on the viewpoint that practice and knowledge are inseparable and that "it does not make sense to talk about either knowledge or practice with- out the other" (Orlikowski, 2002).

An acceptance of the mutual dependence of practice and context leads us to understand that we must change the way we think about software process. We have, from the beginning, been presented with many different development models and methodologies, each pre-architected with a specific environment and/or set of objectives in mind (Avison and Pries-Heje, 2008). Assumptions are often implicit and so unavailable for serious reflection. This approach of proposing generic process 'solutions' can no longer be justified. In addition, the intertwining of practice and context means that context alters as a result of practice, and so our viewing of 'context' as some- thing fixed is also no longer helpful. An inexperienced developer who participates in a series of formal design reviews is changed by the experience. Rather than continuing to propose and justify sets of defined practices within fixed contexts, we must follow other disciplines and turn our focus towards understanding more deeply the relationships between practice and context. Many researchers have sought such under- standing, mainly by examining practice efficacy in a specific situation. For a comprehensive overview of this research, see (McLeod and MacDonell, 2011).

We suggest that such investigations, although interesting, have limited utility because of the large number of possible contextual factors. Without a theoretical model of the problem space, our investigations must necessarily remain fragmented. In this paper, we address this gap by first abstracting the *problem space* and then aiming to understand the relationships between practices and our abstraction. Our model is based on observations made by Orlikowski during a study of distributed development (Orlikowski, 2002) (see section 2). We propose a six-dimensional model, with dimensions *organisational drivers (why)*, *space and time (where)*, *culture (who)*, *product life-cycle stage (when)*, *product constraints (what)* and *engagement constraints (how)*. We test our model by analysing a reported study into the causes of overscoping in a large organisation, and explore possible explanations for findings.

The contributions of this paper are a novel approach to understanding situated software practices and an abstraction of the software problem space based on work in other disciplines. In section 2, we overview studies aimed at problem space abstraction. In section 3, we present our proposed model. In section 4 we apply it to understand an implementation study from the literature and in section 5 we discuss findings. In section 6, we summarise the paper and discuss limitations and future work.

## 2. ABSTRACTING THE PROBLEM SPACE

In this section, we overview research aimed at categorising the problem space along various dimensions.

Avison and Pries-Heje aimed to support selection of a suitable methodology that is project-specific (Avison and Pries-Heje, 2008). For a given project, the authors plotted position along each of eight dimensions on a radar graph and inferred an appropriate methodology from the shape of the plotted graph.

While we support the intent to understanding project-space in this way, we see two limitations in the work. First, the abstracted categories are based on inputs from a single organisation and so are inevitably scoped to the operating space for the organisation. This means that, although key ideas such as distance, quality, and culture are included, some important contexts are missing. One example is the need to consider the business goals of the organsation when selecting practices (Wang et al., 2012). Lepmets et al. suggest that processes "are part of a larger organizational system and therefore cannot be tinkered with in isolation" (Lepmets et al., 2012). A second limitation is that the abstraction is based at the level of the *project*. In the realm of software development this has implications of spanning requirements determination through to product delivery and we suggest that this scope is not sufficiently broad. For example, the most important practices recommended for new product development are at 'the leading edge' i.e. involve issues of strategy and product determination, and these occur before a 'project' to develop the new product is commenced. Another example involves the 'software-as-a-service (SaaS)' delivery paradigm. Here we note that the emphasis changes from a 'developer driven' to a 'customer-driven' environment, where the on-going relationship between development group and customer becomes key (Stuckenberg and Heinzl, 2010). Again, this is not part of a 'project' as we generally understand it.

Clarke and O'Connor propose a reference frame- work for situational factors affecting software development (Clarke and O'Connor, 2012). One aim is to "develop a profile of the situational characteristics of a software development setting" and to use this to support process definition and optimisation. The framework includes eight classifications: *Personnel*, *Requirements*, *Application*, *Technology*, *Organisation*, *Operation*, *Management* and *Business*, further divided into 44 factors. Our critique of this approach is that it remains 'factors-based'. Although factors are grouped into classifications, there is no meaning that helps us understand relevance. For example, the factor 'Cohesion' represents a number of different kinds of 'meaning', including "team members who have not worked for you", "ability to work with uncertain objectives" and "team geographically distant". These three meanings can be viewed as quite different.

Kruchten presents a contextual model based on experience for situating agile practices. The aim of the model is to "guide the adoption and adaptation of agile development practices", particulary in contexts that are "outside of the agile sweet spot" (Kruchten, 2011). The model is interesting to those committed to an assumption of 'agile practices as basis but may be adapted'. However, it is situated in the solution space of agile projects. Wang notes that the need to align practices with business goals is a limitation of agile practices which "do not generally concern themselves with the surrounding business context in which the software development is taking place" (Wang et al., 2012). As for the Avison and Pries-Heje study above, this model excludes some important contexts and has narrower scope than we require.

Orlikowski carried out an exploration of a globally-dispersed, multinational product development organisation, and observed a number of boundaries that served to shape and challenge the distributed product development (Orlikowski, 2002). These boundaries are *temporal* (multiple time zones), *geographical* (multiple global locations), *social* (many participants engaged in joint development work), *cultural* (multiple nationalities), *historical* (different product versions), *technical* (complex system, various infrastructures, variety of standards) and *political* (different interests, local versus global priorities). In Orlikowski's research, the *problem space* is de- scribed. However, the work does not relate specifically to software organisations and the criticism of 'single organisation' remains. Aspects such as business goals and the ongoing relationship between customer(s) and development group are not included.

Zachman created a 2-dimensional framework for describing an enterprise architecture (Zachman, 2009). The first framework dimension comprises the columns *Why*, *Who*, *What*, *Where*, *When* and *How* and the second dimension contains a number of perspectives on the organisation. Although the *meanings* of the first dimension columns are not useful to our re- search (for example, the meaning of *What* as a list or model of artifacts does not support practice categorisation), we recognise the column categories as providing a potentially sound, orthogonal basis for our investigation. Zachman's notion of differing perspectives on the organisation is not relevant for our study.

| Aspect | Examples |
|---|---|
| Where | Physical distance: temporal, locational. |
| Who | Consistency in world views: affected by nationalities, culture, team structure, power structures, etc. |
| When | Life-cycle stage of the situated product. |
| What | Product-related constraints: affected by standards, external product interfaces, required quality, etc. |
| How | Engagement constraints: affected by client delivery expectations, expected involvement, etc. |
| Why | Organisational drivers: result in strategies that cause constraints in other 5 dimensions. |

Figure 1: Model dimensions.

## 3. DIMENSIONAL MODEL OF THE PROBLEM SPACE

The overall goal for our research is to understand the relationships between practices and context in order that we might advise organisations on practices that are appropriate for their particular situation. The current goal is to *establish a model based on the soft-ware problem space that will support investigations into situated practices.*

As preliminary to creating a workable model, we studied the literature in other disciplines, for example, management, organisational science and product development. The outcome is an abstraction based on the works of Orlikowski (Orlikowski, 2002) and Zachman (Zachman, 2009). The abstraction has six dimensions *Why, Who, What, Where, When* and *How*, each dimension representing a key aspect of the problem space. In Figure 1, we overview the meanings of these aspects along with some examples.

**Space and Time (Where).** How stakeholders are separated in space and time will place constraints upon practices relating to communication and co-ordination. The separation might affect any team interface and so this dimension has many possible scenarios. Some examples are remote clients, outsourcing testing and non-colocated teams.

**Culture (Who).** The world-views of stakeholders will influence which practices are likely to be successful. This dimension is manifested in how people are structured to carry out work. Issues of power, language and expectations about 'how things work' will be relevant. A group within a larger organisation may be constrained to adopt the processes of the larger group. A management team and development team will probably have different world-views. A group that likes and expects change might want to experiment with new technologies. A team of developers that works on several projects at the same time differs culturally from a team that is dedicated to a single project.

**Product Life-cycle Stage (When).** A situated product moves through a tailored life-cycle that includes, for example, initial creation, adolescence, maturity and retirement. It is likely that the practices that affect outcomes will be different at various points in the cycle (Kirk and MacDonell, 2009). For example, much of the new product development (NPD) literature indicates that practices such as *clarification of NPD goals, identification of value proposition, cross-functional teams and go/no-go gates* are more important than 'in-development' practices for successful outcomes. Later in the cycle, when the product is in use and there are many requests for new functionality from different sources, change management practices become key. Still later, when the product is nearing retirement, it is likely that practices to keep 'important' customers happy may be most important, during a transition to a newer, more relevant product. MacCormack et al. discuss three life-stage related contexts: new product start-up, best approached by development practices that support emergence; product growth, which requires an agile approach for man- aging rapid product evolution; and product maturity, which requires efficient processes that reduce costs. (MacCormack et al., 2012). Furneaux and Wade discuss product discontinuance and provide a model to support the formation of clear dis- continuance intentions as a prelude to decision- making (Furneaux and Wade, 2010).

**Product Constraints (What).** The nature of the software product may require consideration of standards, product external interfaces and quality expectations. Software for the chemical industry may require adherence to specific process and/or product standards. Embedded software, middle- ware and stand-alone applications involve different kinds of product interface and possibly different practices relating to product integration. Soft- ware for medical equipment may require formal quality practices to be in place.

**Engagement Constraints (How).** The demographic of the receivers of the software product may in- fluence product specification and delivery mechanisms. Delivering custom software to a single customer probably indicates practices such as evolutionary specification, many deliveries, prototyping and customer involvement. Deploying a telephony middleware product to a vertical market with different feature expectations may re- quire a well-defined roadmap, product-line approach, strong customer relationships, effective pre-delivery testing and a defined and infrequent delivery cycle (as the receiving organisations may need to ratify middleware within their own systems). Deploying to many probably requires greater focus on requirements and change management. An early adopter market may expect fast delivery with frequent updates.

**Organisational Drivers (Why).** The key strategic goal of the producer organisation will affect strategic decisions which in turn may affect any of the above dimensions. A goal of 'corner the market' may be addressed with a strategy of new product innovation (*when*), fast delivery (*how*) and involving customers in development (*who*). A goal of 'be a great employer' infers a focus on human-related practices, and may result in a decision to adopt new process methodologies as they arise to support a strategy of 'always working on the leading edge of technology' (*who*).

| Dim | Identified root causes | Contexts |
|---|---|---|
| Why | RT scoping too technology-biased. Lack of unified priority. | Company management lacks clear business strategy and vision. |
| Where | Low understanding of each other's viewpoint around cost. Developers unaware of importance of work. No mutual understanding of requirements. Responsibility for scope definition unclear. | DTs physically separate. RTs and DTs physically separate. RTs and DTs physically separate. RTs and DTs physically separate. |
| Who | Developer unavailability in early stages. Developers unaware of importance of work. No mutual understanding of requirements. RTs and DTs disagree about need for early, detailed specs. RT scoping too technology-biased. Responsibility for scope definition unclear. | Developers carry out maintenance in parallel. Structured as RTs and DTs. Structured as RTs and DTs. Structured as RTs and DTs. Structured as management team and RTs. Structured as RTs and DTs. |
| When | Inflow from multiple sources. | Fast evolving platform |
| What | | |
| How | | |

Figure 2: Mapping of root causes of overscoping.

A goal of 'go global' may exclude practices that require heavy involvement of end-users.

## 4. THE CASE STUDY

A model effectively represents a theory and is useful only if it helps explain existing observed phenomena or can be used to predict. In this section, we discuss a study that describes issues with process implementation and illustrate the use of the model to support explanations for the identified issues.

Bjarnason, Wnuk and Regnell investigate the causes of overscoping in large, market-driven organisations by carrying out semi-structured interviews in a case company and then validating results via a questionnaire (Bjarnason et al., 2012). The case company develops embedded systems for a global market and implements a product line approach, where products reuse the common platform's functionality and qualities. Major releases typically contain 60-80 new features, implemented as 20-25 projects with separate requirements teams (RTs) and development teams (DTs) containing 40-80 developers. The organisation was in process of implementing a more agile approach and the research was aimed at understanding the differences between the two approaches.

The authors identify six causes of overscoping and then carry out a root cause analysis on the identified causes. In Figure 2, we show the identified 'root causes' along with the contexts that were found by the authors to be associated with each root cause.

**Inflow from Multiple Sources.** The authors state that a "large, uncontrollable inflow of requirements has the potential to cause overscoping when not managed and balanced against the amount of available capacity". The context is an evolving platform, with each platform release involving "typically around 60-80 new features". This situation is typical of the software product life-stage, adolescence, as there is a need to satisfy in a timely manner many requests from disparate sources (Kirk and MacDonell, 2009). We thus map to the When dimension.

**Lack of Overview of Available Resources.** The root cause was a *failure of the DTs to under- stand each other's viewpoint around cost*, caused by "communication gaps within the software unit and between the DTs" as each DT focussed on local needs. It is unclear to what extent physical separation of the DTs (*Where*) and a difference in world-view between DTs (*Who*) affected the lack of communication. Because the same kinds of role are involved, we assume the key aspect is physical separation and map to *Where*.

**Low Development Team Involvement in Early Stages.** The authors identified two root causes. The first is developer unavailability as developers work on other maintenance projects. The relevant dimension is *Who*. The second concerns gaps in communication between the requirements units and development units, leading to the developers being unaware of the importance of the work. As for the previous item, we do not know if this lack of communication is caused by physical or cultural distance. As it is likely that the different kinds of unit have different world views, we map this root cause to both *Where* and *Who*.

**Requirements not agreed with Development Team.**

A significant root cause was found to be low mutual understanding resulting from communication gaps between different kinds of role and state that these resulted from "physical and organizational distances". We map to *Where* and *Who*.

**Upfront detailed Requirements Specification.** Those involved in requirements specifications reported no issues with early, detailed specification while those downstream had the opposite viewpoint. We map as cultural (*Who*).

**Unclear Vision of Overall Goal.** The RT leaders reported that a lack of clear business strategy and vision "resulted in proposing a project scope from a pure technology standpoint". We are unclear whether the context is one of management failure to clarify business strategy (*Why*) or failure to share strategy with the RTs, causing a cultural disconnect (*Who*), and so map to both. The authors also found weak scope coordination be- tween functional areas resulted in a project scope largely defined by the DTs (instead of the RTs). We again map to both *Where* and *Who*.

| Dim | Identified root causes | Contexts |
|---|---|---|
| Why | RT scoping too technology-biased. Lack of unified priority. | Company management lacks clear business strategy and vision. |
| Where | Low understanding of each other's viewpoint around cost. ~~Developers unaware of importance of work.~~ ~~No mutual understanding of requirements.~~ ~~Responsibility for scope definition unclear.~~ | DTs physically separate. ~~RTs and DTs physically separate.~~ ~~RTs and DTs physically separate.~~ ~~RTs and DTs physically separate.~~ <u>BUs and SUs physically separate.</u> |
| Who | ~~Developer unavailability in early stages.~~ ~~Developers unaware of importance of work.~~ ~~No mutual understanding of requirements.~~ ~~RTs and DTs disagree about need for early, detailed specs.~~ ~~RT scoping too technology-biased.~~ ~~Responsibility for scope definition unclear.~~ | ~~Developers carry out maintenance in parallel.~~ ~~Structured as RTs and DTs.~~ ~~Structured as RTs and DTs.~~ ~~Structured as RTs and DTs.~~ ~~Structured as management team and RTs.~~ ~~Structured as RTs and Dts.~~ <u>Structured as BUs and SUs.</u> <u>DT includes business representative.</u> <u>Separate planning group.</u> <u>Structured as management team and BUs.</u> |
| When | Inflow from multiple sources. | Fast evolving platform |
| What | | |
| How | | |

Figure 3: Profile after agile implementation.

From Figure 2, we observe that, from the perspective of our model, many of the causes of overscoping in this case are associated with physical and cultural separation of individuals with different roles. Two root causes are associated with organisational failure to clarify business strategy and one with the life-cycle stage of the situated product. We now examine changes introduced with the agile implementation.

1. Requirements responsibility transferred from requirements unit into business unit (BU) and soft-ware unit (SU). BU gathers and prioritises features and places in prioritised list. SU estimates effort and delivery date.

2. Cross-functional development teams created for a feature include a customer proxy (from BU) and produce detailed requirements, implement and test the feature. Iteratively refine requirements, effort and time-frame estimations.

3. User stories and acceptance criteria produced.

4. Planning and resource allocation in one plan.

We analyse with the support of our model and show findings in Figure 3.

- The requirements process described in item 1 above has established responsibility for scope definition. We 'remove' this root cause from *Where* and *Who*. It is likely that BUs will scope in a less technology-biased way, and so we remove this root cause from *Who*. We note that we have a new context i.e. a structure of BU and SU. We add this to *Where* and *Who* by underlining.

- The creation of cross-functional teams has removed the physical and cultural distance between requirements engineers and developers. We would expect that root causes addressed would include 'developer awareness of importance', 'mutual understanding of requirements' and the 'disagreement about the need for early, detailed specs'. The latter is not necessarily the case as including a customer proxy on the team doesn't guarantee agreement about specification mechanisms. We remove the first two causes. However, we note the team now includes a customer proxy from the BU. It is likely that this new team member has different viewpoints from existing members. We add this in Who.

- The creation of user stories and acceptance criteria for a feature addresses the disagreement about the need for early, detailed specifications. We re-move this cause.

- An overall approach to planning and resource allocation addresses 'developer unavailability in early stages' (Who) and changes the context of parallel development. However, we are unclear about who owns this overall plan and a possible change in structure is involved, for example, a planning group. As we need to understand all possible effects of the new implementation, for completeness, we add to Who.

Examining Figure 3, we observe that most identified root causes have been addressed. However, some remain and some are new. The issue of technology-biased scoping was associated with a context of lack of management business strategy and a cultural disconnect between management and requirements teams. Muddy management vision pre-change will remain muddy management vision post-change. The issue of DTs focusing on their own concerns probably remains and it is unclear whether the introduced changes will address all issues arising from this. The inflow from multiple sources is unchanged and may be addressed by the new requirements process. How-ever, this is not certain, as there is no reason to believe that the 'champions' within the BUs of the various requirements sources will always agree about scope and priority, particularly if vision is unclear. The new change management process may not be entirely effective. In addition, the move to agile has introduced four cultural changes, each of which may be a source of problems. For example, the different world-views of business and development team members may cause misunderstanding and conflict.

Our analysis suggests that the change to a more agile approach will mitigate issues of over-scoping, but does not address all root causes and introduces some new possible issues. This analysis is consistent with the findings of the authors, who found that "over- scoping was still a challenge for the case company". Respondents reported a need for "clearer company vision" and this is also consistent with our analysis.

## 5. DISCUSSION

Our analysis of the Case based on our model exposed some post-change contexts that suggest some over- scoping issues might remain after process change. In- deed, we identified some new contexts in *Where* and *Who* that might lead to new problems. We, of course, do not know if there is consistency between our ex- planation and remaining issues, but suggest that the use of our model has provided a sound starting point for further investigations.

We also note the importance of understanding cause at a deeper level than is perhaps accepted. For example, identifying a problem of 'poor communication' is only a first step. The deeper cause, for ex- ample, physical or cultural separation must be under- stood prior to improvement initiatives, if these initiatives are to be effective. Use of the proposed model provides a mechanism for such understanding. For example, in the Case, the issue of technology-biased scoping may have been an issue of lack of management strategy or may have been an issue of organisation structure (cultural). We cannot assume the problem will go away when we form cross-functional teams. From Figure 3, we saw that a) if the is- sue was associated with management strategy, cross- functional teams made no difference and b) if organisation structure, the introduction of cross-functional teams introduces a new structure which may itself cause new issues to emerge.

One important aspect of having a model from which to analyse is the ability to predict what may be unsuccessful in a process improvement initiative. For example, Hoda et al. report on the adaptation of agile practices in a study involving 40 practitioners from 16 software development organisations in New Zealand (Hoda et al., 2010). One problem identified concerned the need for detailed documentation to com- ply with certification requirements. Some developers perceived documentation as being anti-agile, causing problems of internal team clashes. Had an extensive analysis of context been carried out in advance, the conflict between an absolute requirement for documentation (*What*) and the evangelism of some team members (*Who*) may have been exposed in advance and mitigation steps taken.

Our approach of establishing a model of the problem space enables us to now ask focussed research questions with a view to building up a body of knowledge concerning suitable practices in a given con- text. For example, if our question is "What practices are suitable for ensuring correct functionality during design?", we must further focus the question with respect to our dimensional model, by considering business objectives, physical separation of stake- holders, cultural considerations, product life-cycle stage, product-related constraints and engagement-related constraints.

## 6. SUMMARY, LIMITATIONS AND FUTURE WORK

We have postulated the need for a holistic, integrated approach to investigating context-related best practice and have proposed an approach based on problem- space contexts. The model has six dimensions: organisational drivers (why), space and time (where), culture (who), product life-cycle stage (when), product constraints (what) and engagement constraints (how). We have tested our model by using it to analyse a re- ported implementation study and have suggested explanations for findings. Our contributions are a novel approach along with a model for understanding situated software practices.

The major limitation for this research is that both model creation and validation have been carried out by the same people with resulting concerns of subjectivity in interpretation of the validation study. As mitigation, we have aimed to describe our reasoning as fully as possible given space constraints.

Our vision is for a model that is 'a minimum spanning set' for the problem space i.e. dimensions are non-overlapping and cover the space. We will next map known factors onto our model to test for this. Other future efforts will include further exploration of the effectiveness of our model and studies into how the 'context blueprint' changes as specific practices are implemented.